\documentclass[a4paper,fleqn]{cas-sc}

\usepackage[numbers]{natbib}

\usepackage{upgreek}

\def\tsc#1{\csdef{#1}{\textsc{\lowercase{#1}}\xspace}}
\tsc{WGM}
\tsc{QE}
\tsc{EP}
\tsc{PMS}
\tsc{BEC}
\tsc{DE}

\begin{document}

\title [mode = title]{Radiation reaction and the acceleration-dependent mass increase of a charged sphere undergoing uniform acceleration}

\author[1]{Teyoun Kang}

\address[1]{School of Natural Science, UNIST, 50 UNIST-gil, Ulju-gun, Ulsan, 44919, Korea}

\author[2]{Adam Noble}

\address[2]{Department of Physics, SUPA and University of Strathclyde, Glasgow G4 0NG, United Kingdom}

\author[2]{Samuel R. Yoffe}

\author[2]{Dino A. Jaroszynski}

\author[1]{Min Sup Hur}
\cormark[1]

\cortext[cor1]{Corresponding author: mshur@unist.ac.kr.}

\begin{abstract}
Photon emission from a uniformly accelerated charge is among the most mysterious physical phenomena. Theories based on the Lorentz-Abraham-Dirac equation mostly conclude that a uniformly accelerated point charge cannot feel radiation reaction. Such a conclusion suggests that the origin of the photon energy is unclear. In this paper, we determine the self-force of a uniformly accelerated \textit{charged sphere} using the Lorentz force equation, with an assumption that the sphere is Lorentz-contracted during the acceleration. For large acceleration, the calculated self-force converges to the radiation reaction (given by the Larmor formula) via a new factor $\gamma_{a}$, which describes an acceleration-dependent increase in the effective mass. This increased mass makes it harder to accelerate the particle (compared to a point-charge), which means more energy should be provided to the particle in order to get the expected acceleration. This extra energy can be interpreted as the origin of the photon energy.
\end{abstract}

\begin{keywords}
radiation reaction \sep uniform acceleration \sep electrodynamics 
\end{keywords}

\maketitle

\section{Introduction}

Since electromagnetic fields were found to carry momentum, it has been generally agreed that a radiating charge must feel radiation reaction (RR) \cite{Burton2014}. Several theoretical models have been suggested to explain RR in the framework of classical electrodynamics, but the RR problem has not been completely solved. The early theories of Lorentz, Abraham and Dirac (LAD), based on point-charges, have unphysical solutions with unbounded acceleration (`runaway solutions') and causality violation (`preacceleration'). Such problems were resolved in a number of modified theories \cite{Landau1975,Ford1991,Burton2014,Capdessus2016}, but none of these models is free of difficulties, and many aspects of RR remain obscure. It was even proposed that Maxwell's equations need to be modified \cite{Rohrlich1997,Gratus2015}. However, it is generally accepted that the problems stem from the incompatibility of classical electrodynamics with the notion of a point charge.

The first hints that an extended particle might resolve the issues of RR came from Eliezer \cite{Eliezer1948}, who found that some (unspecified) charge distribution could be chosen to eliminate the higher order derivatives that were responsible for the unphysical effects. Further support came from Ford and O'Connell \cite{Ford1991}, who obtained the same equation from (the classical limit of) a quantum Langevin equation for a particle with a given form factor. Levine, Moniz and Sharp  have shown more generally that for a charged shell with radius above some threshold preacceleration and runaways are eliminated \cite{Levine1977}. All these considerations were restricted to nonrelativistic dynamics.

Another area of confusion is the radiation reaction of a uniformly accelerated charge. From the Larmor formula, accelerated charges always emit photons, and uniform acceleration should be no exception \cite{Franklin2014}. However, most theories of RR predict that a uniformly accelerated particle feels no friction from radiation \cite{Fulton1960,Boulware1980,Eriksen2000}. In other words, a steady rate of photon emission does not affect the momentum of a particle. This unintuitive conclusion has caused considerable debate in the literature.

Radiation reaction is not just a problem of classical electrodynamics. Self-interaction effects, including RR, occur also in quantum electrodynamics (QED) \cite{Higuchi2006,Ilderton2013}, and make formulating (far less solving) the Schr\"{o}dinger-type equation far from trivial, as RR does to the Lorentz force \cite{Bild2019}. In this regard, the classical RR problem is still worth investigating because it could give insights into the analogous problem of QED. Moreover, owing to the construction of ultra-intense laser facilities \cite{Mourou2019}, the need to understand RR has recently become more urgent. These facilities can generate lasers of $10^{24}~\mathrm{W}/\mathrm{cm}^{2}$ intensity, and in this regime RR is expected not only to be observable, but to fundamentally change the nature of laser-matter interactions \cite{Blackburn2020}. Accordingly, a rigorous but tractable RR theory which can explain the upcoming observations is now required by experimenters. If a classically obtained theory meets this requirement, it will bridge the gap between classical and quantum electrodynamics.

In this paper we present a new calculation of the classical self-force on a uniformly accelerated charge. The particle model we adopt is not Dirac's model (point charge), but that of Lorentz (charged sphere). Such a choice is made to avoid the inconsistencies of the point-charge that lead to unphysical behaviour. A similar work has been reported in Ref.~\cite{Steane2015}, where the self-force on a \textit{homogeneously} charged sphere under uniform acceleration was calculated. In contrast, in the model used in this paper, we assume that the sphere can be Lorentz-contracted and it bears an image of a point-charge on its surface so that the sphere and the corresponding point-charge are indistinguishable (when observed from outside). Remarkably, our new calculation yields the radiation reaction of a uniformly accelerated charge. Moreover, we show the self-force can be formulated in a simple closed form. Furthermore, we find that the self-force leads to an acceleration-dependent increase in the effective mass of the particle. This increased (effective) mass can be interpreted as the energy source of photons, as it returns to the original mass after the radiation is over.

Throughout this paper, we employ the Einstein summation convention, and use the Minkowski metric $g_{\mu\nu}$ with signature $\left\{-+++\right\}$. SI units are used for all of the equations.
 
\section{Uniformly accelerated charge}

Denoting the position and velocity of a particle by $\boldsymbol{\upchi}$ and $\mathbf{u}$ ($=\mathrm{d}\boldsymbol{\upchi}/\mathrm{d}\tau$, with the proper time $\tau$), Dirac's radiation reaction is given by
\begin{equation}
K_{\mathrm{rad}}^{\mu} = \displaystyle\frac{\mu_{0}q^{2}}{6\pi c} \left(\displaystyle\frac{\mathrm{d}^{2}u^{\mu}}{\mathrm{d}\tau^{2}} - \displaystyle\frac{u^{\mu}}{c^{2}} \displaystyle\frac{\mathrm{d}u_{\nu}}{\mathrm{d}\tau} \displaystyle\frac{\mathrm{d}u^{\nu}}{\mathrm{d}\tau}\right) , \label{eq:Dirac}
\end{equation}
where $\mu_{0} = 4\pi \times 10^{-7} \;\mathrm{kg}\cdot\mathrm{m}\cdot\mathrm{C}^{-2}$ is the vacuum permeability, $q$ is the particle's charge, and $c$ is the speed of light. This force was derived by the Taylor series expansion of the electromagnetic field $F^{\mu\nu}$ near a point charge \cite{Dirac1938}, and it can be considered as RR because it includes a friction term with magnitude equal to the Larmor formula,
\begin{equation}
P_{\mathrm{rad}} = \displaystyle\frac{\mu_{0}q^{2}}{6\pi c} \displaystyle\frac{\mathrm{d}u_{\nu}}{\mathrm{d}\tau} \displaystyle\frac{\mathrm{d}u^{\nu}}{\mathrm{d}\tau} .
\end{equation}
This formula represents the total power of the radiation emitted by an accelerated point-charge. The problem is that if a point-charge is hyperbolically (or uniformly) accelerated, Dirac's RR vanishes while the radiated power does not ($P_{\mathrm{rad}} > 0$); the hyperbolic motion is given by
\begin{equation}
\boldsymbol{\upchi}\left(t\right) = \sqrt{z_{c}^{2} + c^{2}t^{2}} \hat{\mathbf{z}} , \label{eq:hyperbolic}
\end{equation}
where $z_{c}$ is an arbitrary constant and $\hat{\mathbf{z}}$ is the unit vector in the direction of the constant acceleration. Substitution of Eq.~(\ref{eq:hyperbolic}) into Eq.~(\ref{eq:Dirac}) leads to $K_{\mathrm{rad}}^{\mu} = 0$. This apparent contradiction is usually attributed to the reversible exchange of energy with the particle's bound field \cite{Steane2015b}. However, since a particle with finite size has a finite bound energy, this argument cannot fully explain the origin of the radiation by an extended particle undergoing indefinite hyperbolic motion.

To illuminate this problem, we revisit Lorentz's model of a particle as a charged sphere. A homogeneously charged spherical surface at rest generates the static potential of a point-charge ($=q/4\pi\epsilon_{0}r$) in its exterior, but the field inside the sphere vanishes. In this sense the Lorentz model could be generalized as follows:
\begin{enumerate}
\item
Outside, the sphere generates the field of a point-charge under the same motion.
\item
Inside, the field identically vanishes.
\end{enumerate}
Hence, in this paper we will call a particle satisfying these conditions, a `Lorentz particle (LP).' A Lorentz particle cannot be distinguished from an ideal point-charge (unless the inside is observable). In contrast to the result of Dirac's model (which is based on a point-charge), we will demonstrate that a uniformly accelerated LP gets a recoil by RR, via its charge distribution.

In order to find the charge and current distribution of an LP, we employ the Rindler frame \cite{Eriksen2004} represented by barred coordinates $\bar{x}^{\mu} = \left\{c\bar{t},\bar{x},\bar{y},\bar{z}\right\}$, where
\begin{equation}
\begin{aligned}
c\bar{t} &= z_{c} \tanh^{-1}\left(ct/z\right) , \\
\bar{x} &= x ,\\
\bar{y} &= y ,\\
\bar{z} &= \sqrt{z^{2} - c^{2}t^{2}} .%
\end{aligned}
\end{equation}
In this frame, the (flat) Minkowski metric takes the form
\begin{equation}
\bar{g}_{\mu\nu} = \mathrm{diag}\left\{-\bar{z}^{2}/z_{c}^{2},1,1,1\right\} ,
\end{equation}
and a point-charge satisfying Eq.~(\ref{eq:hyperbolic}) will remain at rest at $\bar{x}^{\mu} = \left\{c\bar{t},0,0,z_{c}\right\}$. Correspondingly, the uniform acceleration becomes a static problem. If we suppose $\bar{g}$ is the determinant of the metric ($=-\bar{z}^{2}/z_{0}^{2}$), Maxwell's equations in the Rindler frame can be written
\begin{equation}
\begin{aligned}
\bar{\partial}_{\nu} \left(\bar{F}^{\mu\nu} \sqrt{-\bar{g}}\right) &= \mu_{0}\bar{j}^{\mu} \sqrt{-\bar{g}} , \\
\bar{\partial}_{\mu}\bar{A}_{\nu} - \bar{\partial}_{\nu}\bar{A}_{\mu} &= \bar{F}_{\mu\nu} ,
\end{aligned} \label{eq:Maxwell}
\end{equation}
and accordingly the point-charge at the position $\bar{x}^{\mu} = \left\{c\bar{t},0,0,z_{c}\right\}$ leads to
\begin{align}
\bar{j}_{0}\left(\bar{x},\bar{y},\bar{z}\right) &= -qc \displaystyle\frac{\bar{z}}{z_{c}} \delta\left(\bar{x}\right) \delta\left(\bar{y}\right) \delta\left(\bar{z}-z_{c}\right) , \\
\bar{A}_{0}\left(\bar{x},\bar{y},\bar{z}\right) &= -\displaystyle\frac{\mu_{0}qc}{4\pi} \displaystyle\frac{1}{z_{c}} \displaystyle\frac{\bar{r}^{2} + z_{c}^{2}}{\sqrt{\left(\bar{r}^{2} + z_{c}^{2}\right)^{2} - 4z_{c}^{2}\bar{z}^{2}}} , \label{eq:pointA}
\end{align}
where $\bar{r} = \sqrt{\bar{x}^{2} + \bar{y}^{2} + \bar{z}^{2}}$, and spatial components of the current $\bar{j}_{\mu}$ and potential $\bar{A}_{\mu}$ vanish. These results exactly correspond to the published solutions \cite{Fulton1960,Eriksen2004}.

On a spherical surface given by
\begin{equation}
\bar{x}^{2} + \bar{y}^{2} + \left(\bar{z} - \sqrt{z_{c}^{2} + R_{c}^{2}}\right)^{2} = R_{c}^{2} ,
\end{equation}
the potential $\bar{A}_{0}$ satisfies
\begin{equation}
\left.\bar{A}_{0}\right|_{\text{on the surface}} = -\displaystyle\frac{\mu_{0}qc}{4\pi} \displaystyle\frac{\sqrt{z_{c}^{2} + R_{c}^{2}}}{z_{c}R_{c}} ,
\end{equation}
which indicates that the potential is constant on the spherical surface where $R_{c}$ is the radius. From this notion, we define coordinates $\left\{\bar{R},\bar{\Theta},\bar{\phi}\right\}$ and potential distribution $\bar{A}_{0}'$, defined as
\begin{equation}
\begin{aligned}
 \bar{R} &\equiv \sqrt{\bar{x}^{2} + \bar{y}^{2} + \left(\bar{z} - \sqrt{z_{c}^{2} + R_{c}^{2}}\right)^{2}} , \\
 \bar{\Theta} &\equiv \tan^{-1}\left(\displaystyle\frac{\sqrt{\bar{x}^{2}+\bar{y}^{2}}}{\bar{z}-\sqrt{z_{c}^{2} + R_{c}^{2}}}\right) , \\
 \bar{\phi} &\equiv \tan^{-1}\left(\displaystyle\frac{\bar{y}}{\bar{x}}\right) , \\
\end{aligned}
\end{equation}
and
\begin{equation}
\bar{A}_{0}'\left(\bar{x},\bar{y},\bar{z}\right) \equiv \begin{cases}
-\displaystyle\frac{\mu_{0}qc}{4\pi} \displaystyle\frac{\sqrt{z_{c}^{2} + R_{c}^{2}}}{z_{c}R_{c}} & \left(\bar{R} \leq R_{c}\right) , \\
\bar{A}_{0}\left(\bar{x},\bar{y},\bar{z}\right) & \left(\bar{R} > R_{c}\right) . \\
\end{cases}
\end{equation}
This potential is equal to Eq.~(\ref{eq:pointA}), i.e. the potential of a point charge, outside the sphere ($\bar{R}>R_{c}$), but constant for $\bar{R} \leq R_{c}$, so that the field vanishes inside the sphere. The charge density $\bar{j}_{0}'$ inducing this potential can be obtained from Maxwell's equation (\ref{eq:Maxwell}):
\begin{align}
\bar{j}_{0}'\left(\bar{x},\bar{y},\bar{z}\right) &= -\displaystyle\frac{1}{\mu_{0}} \displaystyle\frac{\bar{z}}{z_{c}} \bar{\partial}_{\nu} \left[\displaystyle\frac{z_{c}}{\bar{z}} \bar{\partial}^{\nu} \bar{A}_{0}'\left(\bar{x},\bar{y},\bar{z}\right)\right] \nonumber \\
&= -\displaystyle\frac{qc}{4\pi R_{c}^{2}} \displaystyle\frac{z_{c}}{\bar{z}} \delta\left(\bar{R}-R_{c}\right). \label{eq:density}
\end{align}
This charge density completely meets the requirements for LP: outside the sphere ($\bar{R}>R_{c}$) it induces the field of the point charge ($\bar{A}_{0}' = \bar{A}_{0}$), and inside ($\bar{R} \leq R_{c}$), the field vanishes ($\bar{A}_{0}' = \mathrm{const}$). Consequently, external observers cannot distinguish the LP from the point-charge, as they observe an image point-charge on the surface.

\begin{figure}
\centering
\includegraphics[scale=0.4]{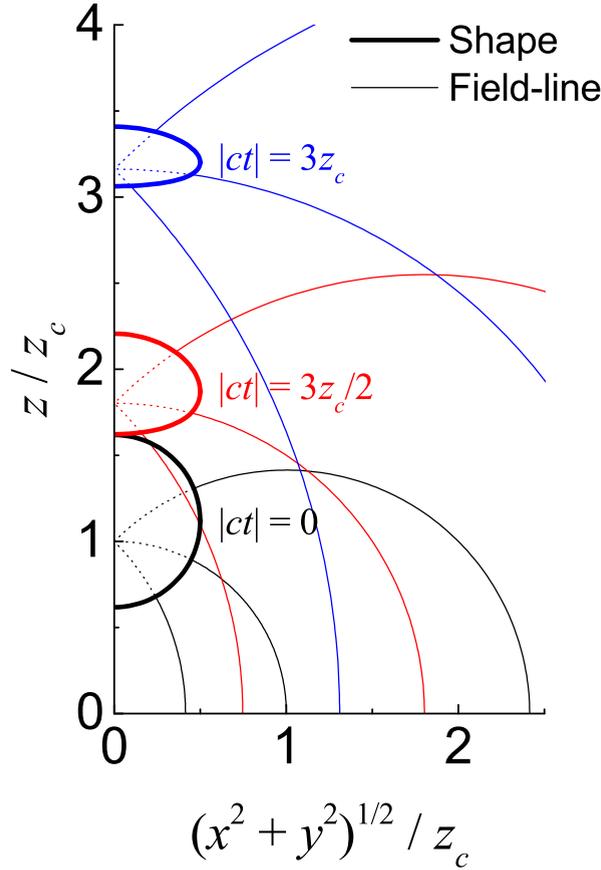}
\caption{Shapes and electric field-lines of a uniformly accelerated LP at three different times, where $R_{c} = z_{c}/2$. The dotted lines represent the field lines which would be generated if the LP were a point-charge.} \label{fig:LP}
\end{figure}

Although the Lorentz particle has a finite volume, we can define its position as the location of the point charge which would generate the same external field as the LP. The position of a uniformly accelerated LP defined in this way satisfies Eq.~(\ref{eq:hyperbolic}). Figure~\ref{fig:LP} shows the LP's positions and actual shapes at three different times observed in the lab frame. As the field lines in Fig.~\ref{fig:LP} indicate, according to Gauss's law, the total charge of the LP is always $q$;
\begin{align}
\int j_{0}' \,\mathrm{d}^{3}r &= \int \bar{j}_{0}' \displaystyle\frac{z_{c}}{\bar{z}} \,\mathrm{d}^{3}\bar{r} , \nonumber \\
&= -\displaystyle\frac{qc}{4\pi R_{c}^{2}} \int \displaystyle\frac{z_{c}^{2}}{\bar{z}^{2}} \delta\left(\bar{R}-R_{c}\right) \mathrm{d}^{3}\bar{r} , \nonumber \\
&= -\displaystyle\frac{qc}{4\pi R_{c}^{2}} \int \displaystyle\frac{z_{c}^{2}\bar{R}^{2}\sin\bar{\Theta} \delta\left(\bar{R}-R_{c}\right)}{\left(\bar{R}\cos\bar{\Theta} + \sqrt{z_{c}^{2}+R_{c}^{2}}\right)^{2}} \,\mathrm{d}\bar{R}\mathrm{d}\bar{\Theta}\mathrm{d}\bar{\phi} , \nonumber \\
&= -qc . \label{eq:charge}
\end{align}


In contrast, we show that the momentum of the LP does not coincide with that of the point-charge. The original expression of the Lorentz force is given by
\begin{equation}
\displaystyle\frac{m}{q} \displaystyle\frac{\mathrm{d}}{\mathrm{d} t} \int \rho \mathbf{U} \,\mathrm{d}^{3}r = \int (\rho \mathbf{E} + \mathbf{J} \times \mathbf{B}) \,\mathrm{d}^{3}r , \label{eq:Lorentz}
\end{equation}
where $m$ is the (unrenormalized) bare mass and $\mathbf{U}$ is relativistic velocity of the charge distribution, defined as $\mathbf{U} \equiv c\mathbf{J} / \sqrt{-j_{\nu}j^{\nu}}$, and we assume the charge and mass distributions are proportional. In the case of the point-charge, the momentum is simply $m\mathbf{u}$ because $\rho = q\delta\left(\mathbf{r}-\boldsymbol{\upchi}\right)$. On the other hand, in case of the LP, the charge is distributed over a finite surface, and accordingly the momentum is determined by the volume integration
\begin{align}
\mathbf{p} &= \displaystyle\frac{m}{q} \int \rho' \mathbf{U}' \,\mathrm{d}^{3}r , \nonumber \\
&= -\displaystyle\frac{m}{qc} \hat{\mathbf{z}} \int j_{0}' U_{3}' \,\mathrm{d}^{3}r .
\end{align}
As the LP remains at rest in the Rindler frame, we obtain $\bar{U}_{\mu}' = \left\{-c\bar{z}/z_{c},0,0,0\right\}$, and the Lorentz boost yields
\begin{align}
\mathbf{p} &= \displaystyle\frac{m}{qc} \hat{\mathbf{z}} \int \bar{j}_{0}' \bar{U}_{0}' \displaystyle\frac{ct}{\bar{z}} \displaystyle\frac{z_{c}^{2}}{\bar{z}^{2}} \,\mathrm{d}^{3}\bar{r} , \nonumber \\
&= \displaystyle\frac{mc^{2}t}{4\pi R_{c}^{2}} \hat{\mathbf{z}} \int \displaystyle\frac{z_{c}^{2}}{\bar{z}^{3}} \delta\left(\bar{R}-R_{c}\right) \mathrm{d}^{3}\bar{r} , \nonumber \\
&= \displaystyle\frac{mc^{2}t}{4\pi R_{c}^{2}} \hat{\mathbf{z}} \int \displaystyle\frac{z_{c}^{2}\bar{R}^{2}\sin\bar{\Theta} \delta\left(\bar{R}-R_{c}\right)}{\left(\bar{R}\cos\bar{\Theta} + \sqrt{z_{c}^{2}+R_{c}^{2}}\right)^{3}} \,\mathrm{d}\bar{R}\mathrm{d}\bar{\Theta}\mathrm{d}\bar{\phi} , \nonumber \\
&= \displaystyle\frac{mc^{2}t}{z_{c}} \sqrt{1 + \displaystyle\frac{R_{c}^{2}}{z_{c}^{2}}} \hat{\mathbf{z}} .
\end{align}
Variables in this equation can be substituted by the velocity and acceleration using $\mathbf{u} = \hat{\mathbf{z}}c^{2}t/z_{c}$ and $\mathrm{d}\mathbf{u}/\mathrm{d}t = \hat{\mathbf{z}}c^{2}/z_{c}$, leading to the final form of the momentum,
\begin{equation}
\mathbf{p} = m\mathbf{u} \sqrt{1 + \displaystyle\frac{R_{c}^{2}}{c^{4}} \left|\displaystyle\frac{\mathrm{d}\mathbf{u}}{\mathrm{d}t}\right|^{2}} . \label{eq:momentum}
\end{equation}


What is indicated by Eq.~(\ref{eq:momentum}) is interesting: the LP's momentum is always greater than the momentum of the corresponding point-charge. This is the consequence of the change in the shape of the LP every moment as shown in Fig.~\ref{fig:LP}; to be Lorentz-contracted, the rear and front surfaces of the LP need to move faster and slower than the corresponding point-charge, and Eq.~(\ref{eq:density}) indicates that the charge density on the rear surface is higher than that on the front surface ($\left|\bar{j}_{0}'\right|_{\bar{z}<z_{c}} > \left|\bar{j}_{0}'\right|_{\bar{z}>z_{c}}$), which means the average velocity of the LP is greater than $\mathbf{u}$. It leads to the same effect as increasing the mass, though the actual total mass of the LP is the constant $m$, see Eq.~(\ref{eq:charge}) (assuming the mass density of the LP is $m\rho'/q$). The increase is a purely geometrical effect, and might be assumed irrelevant to radiation reaction. However, we now show that the square root term in Eq.~(\ref{eq:momentum}) also appears in the self-force on the LP, and eventually this term yields the radiation reaction.

The self-force is the force applied to the charge by the field produced by the charge itself. Normally, electromagnetic self-force has been believed to include the radiation reaction. The self-force on a point-charge must be regularized as it diverges, leading to ambiguous results. In contrast, according to Eq.~(\ref{eq:Lorentz}) the self-force on the LP can be represented by
\begin{align}
\mathbf{F}_{\mathrm{self}} &= \int (\rho'\mathbf{E}' + \mathbf{J}' \times \mathbf{B}') \,\mathrm{d}^{3}r , \nonumber \\
&= \hat{\mathbf{z}} \int g^{\mu\nu} j_{\mu}' F_{3\nu}' \,\mathrm{d}^{3}r .
\end{align}
In a similar way to the momentum calculation, the integral is calculated as
\begin{align}
\mathbf{F}_{\mathrm{self}} &= -\hat{\mathbf{z}} \int \bar{j}_{0}' \bar{\partial}_{3} \bar{A}_{0}' \displaystyle\frac{z_{c}^{2}}{\bar{z}^{2}} \,\mathrm{d}^{3}\bar{r} , \nonumber \\
&= \displaystyle\frac{\mu_{0}q^{2}c^{2}}{32\pi^{2} R_{c}^{4}} \hat{\mathbf{z}} \int \displaystyle\frac{z_{c}^{4}}{\bar{z}^{4}} \displaystyle\frac{\bar{z} - \sqrt{z_{c}^{2}+R_{c}^{2}}}{R_{c}} \delta\left(\bar{R}-R_{c}\right) \mathrm{d}^{3}\bar{r} , \nonumber \\
&= \displaystyle\frac{\mu_{0}q^{2}c^{2}}{64\pi^{2} R_{c}^{5}} \hat{\mathbf{z}} \int \displaystyle\frac{z_{c}^{4}\bar{R}^{3}\sin\left(2\bar{\Theta}\right) \delta\left(\bar{R}-R_{c}\right)}{\left(\bar{R}\cos\bar{\Theta} + \sqrt{z_{c}^{2}+R_{c}^{2}}\right)^{4}} \,\mathrm{d}\bar{R}\mathrm{d}\bar{\Theta}\mathrm{d}\bar{\phi} , \nonumber \\
&= -\displaystyle\frac{\mu_{0}q^{2}}{6\pi R_{c}} \displaystyle\frac{c^{2}}{z_{c}} \sqrt{1 + \displaystyle\frac{R_{c}^{2}}{z_{c}^{2}}} \hat{\mathbf{z}} .
\end{align}
By the same substitution of $z_{c}$ by acceleration used in the momentum calculation, the self-force is written
\begin{equation}
\mathbf{F}_{\mathrm{self}} = -\displaystyle\frac{\mu_{0}q^{2}}{6\pi R_{c}} \displaystyle\frac{\mathrm{d}\mathbf{u}}{\mathrm{d}t} \sqrt{1 + \displaystyle\frac{R_{c}^{2}}{c^{4}} \left|\displaystyle\frac{\mathrm{d}\mathbf{u}}{\mathrm{d}t}\right|^{2}} . \label{eq:selfforce}
\end{equation}
As mentioned, the self-force also shows the same square root factor as the momentum. This factor tends to unity as the radius $R_{c}$ approaches zero, and the diverging $\mathbf{F}_{\mathrm{self}}$ for a small $R_{c}$ can be treated as the mass renormalization the same way as in Dirac's theory for a point-charge. However, for finite $R_{c}$ an important feature of Eq.~(\ref{eq:selfforce}) is that the renormalized mass depends on the acceleration.

Since the self-force can be absorbed into a mass renormalization it might seem that, like the point-charge, the LP under uniform acceleration does not feel radiation reaction. However, interestingly, if the acceleration is very large Eq.~(\ref{eq:selfforce}) converges to the Larmor formula:
\begin{equation}
\left|\displaystyle\frac{\mathrm{d}\mathbf{u}}{\mathrm{d}t}\right|\gg \displaystyle\frac{c^{2}}{R_{c}} \implies \mathbf{F}_{\mathrm{self}} \simeq -\displaystyle\frac{P_{\mathrm{rad}}}{c} \hat{\mathbf{z}}. \label{eq:reaction}
\end{equation}
The remarkable point is that even if the radius is extremely small, we can obtain the above result if the acceleration is large enough\footnote{Here we ignore quantum effects which may become significant when the proper acceleration approaches that induced by the Sauter-Schwinger field, $E_S=m^2_ec^3/e\hbar$, with $m_e$ the electron mass and $e$ the elementary charge. We discuss limitations imposed by $E_S$ at the end of this Section.}. Thus the self-force of the uniformly accelerated LP includes radiation reaction. Furthermore, while previous theories have been approximated by Taylor series or other approximations \cite{Dirac1938,Landau1975}, Eq.~(\ref{eq:selfforce}) is exact. Therefore, a mathematically accurate equation of motion for a uniformly accelerated LP can be derived from Eq.~(\ref{eq:selfforce}).

To understand how the radiation reaction in Eq.~(\ref{eq:selfforce}) can be associated with the mass increase in (\ref{eq:momentum}), it is useful to write the equation of motion of a constantly accelerating LP. The momentum conservation is given by
\begin{equation}
\displaystyle\frac{\mathrm{d}\mathbf{p}}{\mathrm{d}t} = \mathbf{F}_{\mathrm{self}} + \mathbf{F}_{\mathrm{ext}} . \label{eq:conservation}
\end{equation}
Uniform acceleration of an LP can be generated by a homogeneous electric and vanishing magnetic field, yielding the external force
\begin{align}
\mathbf{F}_{\mathrm{ext}} &= \int (\rho'\mathbf{E}_{\mathrm{ext}} + \mathbf{J}' \times \mathbf{B}_{\mathrm{ext}}) \,\mathrm{d}^{3}r , \nonumber \\
&= -\displaystyle\frac{\mathbf{E}_{\mathrm{ext}}}{c} \int j_{0}' \,\mathrm{d}^{3}r , \nonumber \\
&= q\mathbf{E}_{\mathrm{ext}} . \label{eq:external}
\end{align}
Finally, by substituting Eqs.~(\ref{eq:momentum}), (\ref{eq:selfforce}) and (\ref{eq:external}) in Eq.~(\ref{eq:conservation}), we obtain the equation of motion for the uniformly accelerated LP:
\begin{equation}
\left(m + \displaystyle\frac{\mu_{0}q^{2}}{6\pi R_{c}}\right) \displaystyle\frac{\mathrm{d}\mathbf{u}}{\mathrm{d}t} \sqrt{1 + \displaystyle\frac{R_{c}^{2}}{c^{4}} \left|\displaystyle\frac{\mathrm{d}\mathbf{u}}{\mathrm{d}t}\right|^{2}} = q\mathbf{E}_{\mathrm{ext}} . \label{eq:motion1}
\end{equation}
Following Dirac, we define the renormalized mass as
\begin{equation}
m_{q} \equiv m + \displaystyle\frac{\mu_{0}q^{2}}{6\pi R_{c}} . \label{eq:renormalization}
\end{equation}
Note that, unlike the case of the point-charge, this is a finite mass renormalization, and the last term of Eq.~(\ref{eq:renormalization}) originates from the self-energy of a finite charge distribution. It is also notable that the last term of Eq.~(\ref{eq:renormalization}) has dimensions of mass. The square root term remains as the only difference between Eq.~(\ref{eq:motion1}) and the well-known form of Lorentz force. This term increased the effective mass in Eq.~(\ref{eq:momentum}) and yielded the radiation reaction in Eq.~(\ref{eq:selfforce}). This strongly implies that the mass variation and radiation reaction are closely related phenomena. As the role and form of this term are analogous to the Lorentz factor\footnote{The Lorentz factor can be also given by $\gamma = 1/\sqrt{1-\left|\mathbf{v}/c\right|^{2}}$, where $\mathbf{v} = \mathrm{d}\boldsymbol{\upchi}/\mathrm{d}t = \mathbf{u}/\gamma$.} $\gamma$ ($=\sqrt{1+\left|\mathbf{u}/c\right|^{2}}$), we define a new factor $\gamma_{a}$ as
\begin{equation}
\gamma_{a} \equiv \sqrt{1 + \displaystyle\frac{R_{c}^{2}}{c^{4}} \left|\displaystyle\frac{\mathrm{d}\mathbf{u}}{\mathrm{d}t}\right|^{2}} , \label{eq:gamma}
\end{equation}
leading to
\begin{equation}
\gamma_{a} m_{q} \displaystyle\frac{\mathrm{d}\mathbf{u}}{\mathrm{d}t} = q\mathbf{E}_{\mathrm{ext}} . \label{eq:motion2}
\end{equation}
Indeed, by this factor the particle becomes harder to accelerate, which implies that a stronger uniform field should be applied to the LP in order to get the conventional constant acceleration ($=q\mathbf{E}_{\mathrm{ext}}/m_{q}$). In other words, to get the expected acceleration, more energy should be provided to the particle. This extra energy compensates for the energy loss by radiation, as indicated by Eq.~(\ref{eq:reaction}).

It is interesting to note that the factor $\gamma_{a}$ is {\it a priori} independent of the charge $q$ and mass $m_{q}$, depending only on the radius $R_{c}$, which is undetermined by the theory. A `natural' choice of radius is given by $R_{c}=\mu_{0}q^{2}/6\pi m_{q}$, the minimum radius required to eliminate the problems of runaways and preacceleration \cite{Levine1977}. In the case of electrons, this radius is roughly $2\times 10^{-15} \;\mathrm{m}$, so that even the Sauter-Schwinger electric field $E_S$ would induce a relative mass shift $\gamma_a-1$ of only $10^{-5}$. However, there are a number of conjectures referring to a finite radius for the electron, ranging from $10^{-22} \;\mathrm{m}$ to $2\times 10^{-12} \;\mathrm{m}$ \cite{Dehmelt1988,Compton1919,Lake2015}. In the context of radiation reaction, it has been argued \cite{Moniz1977} that quantum uncertainty should imbue the electron with an effective radius equal to the Compton wavelength, $\lambda_C=h/m_e c\simeq 2\times 10^{-12} \;\mathrm{m}$. In this case, the effective mass increases by 5\% in a field strength $E=0.05 E_S$. For the intensity of a laser pulse, $0.05 E_{S}$ corresponds to $6\times 10^{26} \;\mathrm{W}/\mathrm{cm}^{2}$, which is expected to be attained within ten years \cite{Mourou2019}. This suggests that it may be possible to observe the acceleration-dependent mass increase in the near future, if the electron is an LP, while also explaining why this effect was not observed at recent experiments on radiation reaction \cite{Cole2018,Poder2018}, which operated at intensities of $4\times 10^{20} \;\mathrm{W}/\mathrm{cm}^{2}$.

\section{Conclusion}

In this paper we have determined the self-force of a uniformly accelerated and Lorentz-contracted charged sphere, or `Lorentz particle (LP),' which to an external observer is indistinguishable from a correspondingly moving point-charge, while the field inside vanishes. This self-force, Eq.~(\ref{eq:selfforce}), was found, for extremely strong acceleration, to have magnitude equivalent to the Larmor formula  (see Eq.~(\ref{eq:reaction})). Based on momentum conservation (\ref{eq:conservation}), the equation of motion of the uniformly accelerated LP was obtained as Eq.~(\ref{eq:motion2}), and the new factor $\gamma_{a}$ was found to give an acceleration-dependent increase to the effective mass of the LP.

Despite its simplicity, the LP model overcomes a notable limitation of the point particle: Eq.~(\ref{eq:selfforce}) demonstrates that the uniformly accelerated LP does experience radiation reaction, and Eq.~(\ref{eq:motion2}) reveals how the energy of the radiation originates in the reduced acceleration produced by an external force. It is also important to note that these results are mathematically rigorous. More general results remain unknown, and we will seek further support for the LP model by exploring other trajectories in future work.

\section*{Acknowledgment}

This work is supported by the Basic Research in Science \& Engineering Program through the National Research Foundation (NRF) of Korea (under Grant Nos. NRF-2016R1A5A1013277, NRF-2017M1A7A1A03072766, NRF-2020R1I1A1A01067536, and NRF-2020R1A2C1102236). Adam Noble, Samuel R. Yoffe, and Dino A. Jaroszynski are supported by the UK EPSRC (Grant No. EP/N028694/1) and the EC's Laserlab-Europe H2020 EC-GA No. 871124.




\end{document}